# Topological fractal networks introduced by mixed degree distribution

Liuhua Zou, Wenjiang Pei, Tao Li, Zhenya He

*Department of Radio Engineering, Southeast University, Nanjing 210096, China*

Yiuming Cheung

*Department of Computer Science, Hong Kong Baptist University, Hong Kong, China*

Several fundamental properties of real complex networks, such as the small-world effect, the scale-free degree distribution, and recently discovered topological fractal structure, have presented the possibility of a unique growth mechanism and allow for uncovering universal origins of collective behaviors. However, highly clustered scale-free network, with power-law degree distribution, or small-world network models, with exponential degree distribution, are not self-similarity. We investigate networks growth mechanism of the branching-deactivated geographical attachment preference that learned from certain empirical evidence of social behaviors. It yields high clustering and spectrums of degree distribution ranging from algebraic to exponential, average shortest path length ranging from linear to logarithmic. We observe that the present networks fit well with small-world graphs and scale-free networks in both limit cases (exponential and algebraic degree distribution respectively), obviously lacking self-similar property under a length-scale transformation. Interestingly, we find perfect topological fractal structure emerges by a mixture of both algebraic and exponential degree distributions in a wide range of parameter values. The results present a reliable connection among small-world graphs, scale-free networks and topological fractal networks, and promise a natural way to investigate universal origins of collective behaviors.

## I. INTRODUCTION

Many artificial and natural complex systems are conveniently modeled with a network, where constituent ingredients and interactions are represented with vertices and links, respectively [1-3]. Systems such as the Internet [4,5], the World Wide Web (WWW) [6], social networks [7], food webs [8], and biological networks [9,10] etc. can be represented as a graph. Strikingly, many of these networks have complex topological properties and dynamical features that cannot be accounted for by classical graph modeling [11]. Recent studies indicate that the realistic networks exhibit some common topological features by a short average distance as random networks, large clustering as regular lattices (small-world effect) [12], a power-law degree distribution (scale-free property) [13], and hierarchical modularity [14,15]. More recently, the emergence of self-similarity in complex networks [16], widely believed as the fractal small-world dichotomy in previous studies [12,17], raises the fundamental question of networks evolution. In this letter, we focus on acceptably social behaviors forced growing network mechanism for a profound view on understanding such common features of realistic complex systems.

The recent discovery of fractal scaling and topological self-similarity in several real networks suggests a common self-organization dynamics [16]. Fractal scaling stands for the scaling relation $N_B / N \sim \ell_B^{-d_B}$ between the number of boxes $N_B$ needed to tile the entire network and the linear size $\ell_B$ with a finite fractal dimension $d_B$ [16]. However, most of the random network models proposed yet are not fractal. Until very recently, Song et al. present self-similar dynamical evolution of complex networks by the inverse renormalization procedure with all of the properties of the network being invariant under time evolution [18]. It has been shown that the key principle that gives rise to fractal structure of networks is a strong effective disassortativity between the hubs on all length scales. In Ref. [18], Model I produces a scale-free, small world network but without fractal topology, while Model II leads to a scale-free network with a fractal topology but not the small-world effect. They suggest growing fractal small-world and scale-free networks from an indirect way of stochastic combination of Model I and Model II. Goh et al. present an *in silico* model with both fractal scaling and scale-free degree distribution based on the multiplicative branching tree. Note also that this fractal trees are not small world, but by introducing a small number of global shortcuts yields small-world property [19]. Different from these growth mechanisms, the present model can grow fractal small-world and mixed scaling networks naturally.

In the present work, we address the study of growing self-similar scale-free networks from empirical evidence of social behaviors: (1) in citation networks, it has been shown [20,21] that the probability for a paper to obtain a new link (citation) is an increasing function of the number of links the node already has (growth and preferential attachment). Papers cease to receive links, because their contents are outdated or summarized in review papers (aging or deactivation mechanism) [22-25]. (2) each of the authors and papers is assigned a topic, and authors read, cite, produce papers only in their own topics (roughly speaking, geographical restriction or limited information) [23,26-28]. (3) more new research branches will be created with deeper studying in certain field (branching). We present a network model with the geographical attachment preference and branching-deactivation mechanisms. We demonstrate that, in a wide range of adjusting parameter, it account for small world property, power law degree distribution, hierarchical organization, and topological fractal.

## II. THE MODEL

The present model starts from an initial configuration of $N_0$ active, completely connected vertices. Each vertex of the present networks can be in two different states, either active or inactive [24,25]. The growth of the model is shown in Fig. 1.

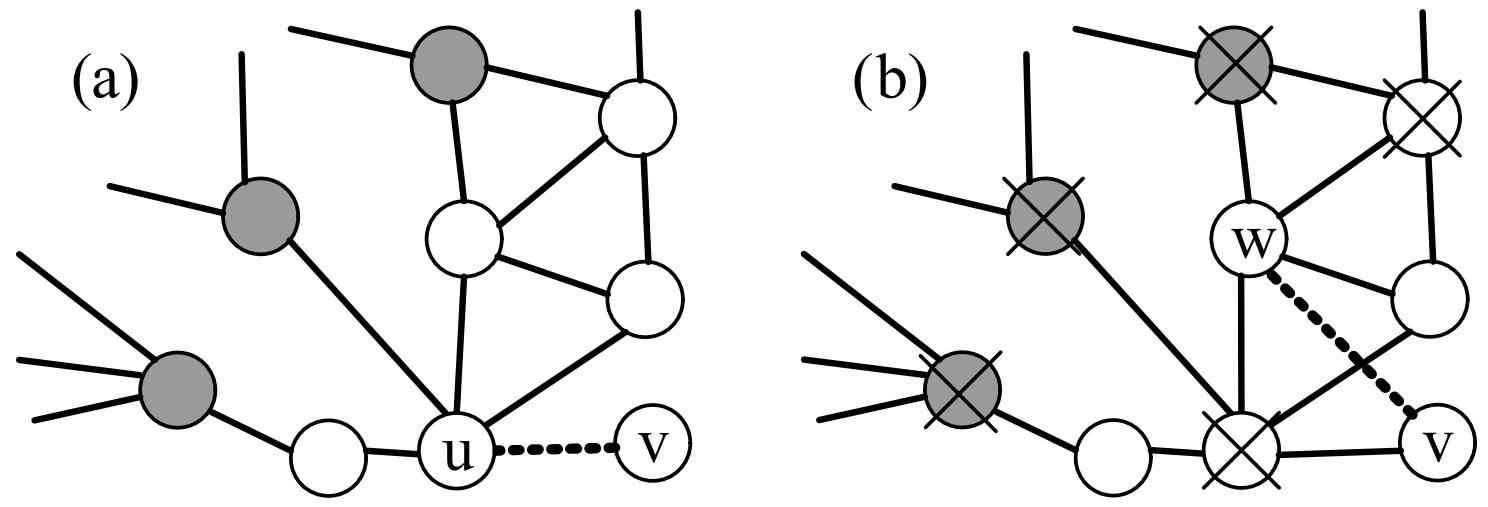

FIG. 1. New vertex creation and geographical attachment. In the new vertex creation step (a) the new node $v$ is created or established by one of the existing active nodes $u$ and links to it. In the geographical attachment step (b) the new vertex $v$ chooses a vertex $w$ in the neighborhood of $u$. It is worth noting that a vertex receives links during the time it is active, and once inactive it will not receive links any longer. Moreover, new vertices have their connectivity influenced by geographical constraints and it forms links locally to the active vertex $u$ and $m-1$ active nearest neighbors of $u$. The filled and unfilled circles stand for inactive and active nodes respectively. $\times$ symbolizes "not allowed to attach to" (either since the vertex is not an active neighbor of the vertex $u$, or that an edge already exists).

At each subsequent discrete time step we grow the network according to the following prescription: (1) A new node $v$ is created or established by one of the existing active nodes $u$ and links to it. (2) The new node $v$ makes another $m-1$ links to $m-1$ active nearest neighbors of $u$. (3) Activate the new node $v$, and then deactivate each active vertex with the probability $P_d(k_i) = \mu \frac{1}{k_i} / \sum_{j \in A} \frac{1}{k_j}$, where $A$ is the set of the currently active nodes. These steps are repeated sequentially, creating a network with a number of nodes $N$ and an average connectivity $\langle k \rangle = 2m$.

## III. TOPOLOGICAL FRACTAL INTRODUCED BY MIXED DEGREE DISTRIBUTION

### A. Degree distribution

The overall degree distribution $N(k)$ can be obtained analytically for the present model, considering the continuous limit of $k$. Let us first derive the degree distribution $p(k,t)$ of the active nodes at time $t$. At time $t$, a node with degree $k = m$ is added to the network, and if it links to a previously existing node $i$, then $k_i \to k_i + 1$. Each preexisting active node is equally likely to be connected to the new node, and therefore the probability that a given preexisting active node has its degree increased by 1 is $\frac{m}{m+(1-\mu)t}$. We define $G(k,t)$ as the number of active nodes with degree $k$ at time $t$. For $k > 0$, the time evolution is determined by the following master equation:

$$G(k+1,t+1) = G(k+1,t)[1-P_d(k+1)][1-\frac{m}{m+(1-\mu)t}] + G(k,t)[1-P_d(k)]\frac{m}{m+(1-\mu)t} + \delta_{km} \quad (1)$$

where $\delta_{km}$ is the Kronecker delta function. The first term on the right-hand side accounts for the process in which an active node with degree $k+1$ at time $t$ is not connected to the new node and still active in the next time step. The second term indicates the process that an active node with degree $k$ at time $t$ is connected to the new node and not deactivated at time $t+1$. The third term represents the new vertex with degree $m$. The degree distribution of active nodes at time $t$ is $p(k,t) = G(k,t)/[m+(1-\mu)t]$. Treating $k$ as continuous, it yields the solution

$$\frac{dp}{dk} = \left(-\frac{1-\mu}{m+1-\mu} + \frac{[\frac{(1-\mu)^2\mu\alpha t}{m+1-\mu} - (1-\mu)\mu\alpha t - m\mu\alpha]k - m\mu\alpha}{(m+1-\mu)k^2 + [m+1-\mu+(1-\mu)\mu\alpha t]k}\right)p(k) \quad (2)$$

where $\alpha = 1/\sum_{i\in A}\frac{1}{k_i}$. In order to determine $\alpha$ over wider ranges for $m$, $\mu$ and $t$ respectively, we make extensive numerical calculations for $m \in [2,10]$, $\mu \in [0,1]$ and $t \in [10^3, 10^4]$. The numerical results as shown in Fig. 2 give that $\alpha \approx \dfrac{1.6m+1}{m+(1-\mu)t}$. The degree distribution of active nodes yields

$$p(k) = be^{-\frac{1-\mu}{m+1-\mu}k} k^{-\frac{\mu(1-\mu)(m+1)(1.6m+1)}{(m+1-\mu)^2(\frac{m}{t}+1-\mu)} - \frac{m\mu\alpha}{m+1-\mu}} \tag{3}$$

with an appropriate normalization constant $b$. We see that $p(k)$ is generally a mixture of both exponential and algebraic distribution.

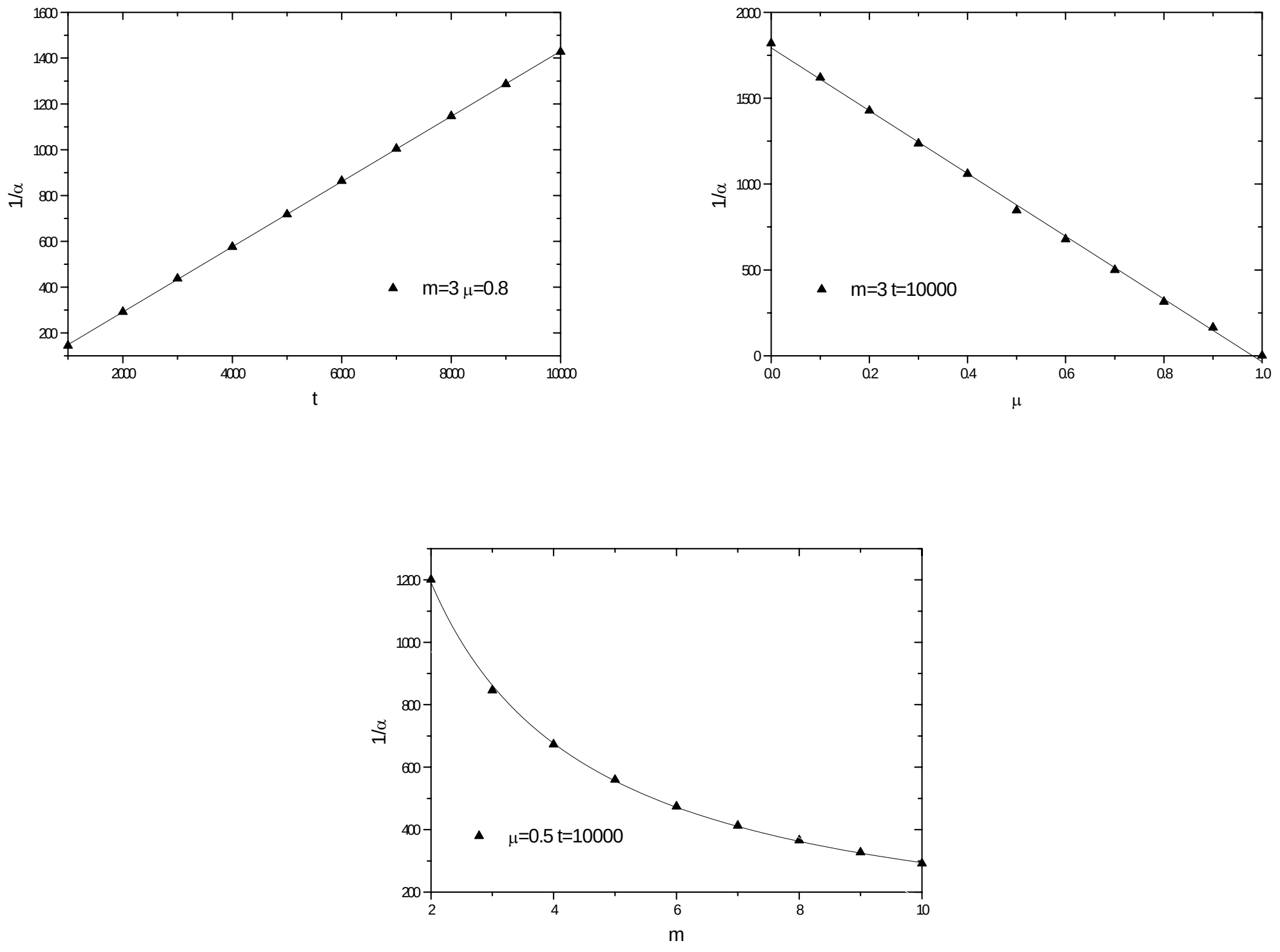

FIG. 2 Numerical calculations to determine $\alpha$ over wider ranges for $m$, $\mu$ and $t$, respectively, where $m \in [2,10]$, $\mu \in [0,1]$ and $t \in [10^3, 10^4]$.

The overall degree distribution $N(k)$ can be calculated by considering both active and inactive nodes. We define $H(k,t)$ as the total number of nodes with degree $k$ in the whole network at time $t$.

$$H(k,t) = G(k,t) + \sum_{t'=1}^{t} P_d(k) G(k,t') \tag{4}$$

Hence, the overall degree distribution $N(k)=\lim_{t\to\infty} H(k,t)/t$ yields

$$N(k)\approx b[(1-\mu)+\frac{\mu(1.6m+1)}{k}]e^{-\frac{1-\mu}{m+1-\mu}k}k^{-\frac{\mu(1-\mu)(m+1)(1.6m+1)}{(m+1-\mu)^2(\frac{m}{t}+1-\mu)}-\frac{m\mu\alpha}{m+1-\mu}} \qquad (5)$$

We see that the overall degree distribution $N(k)$ expresses also a mixed scaling for both algebraic and exponential distributions, which is consistent with many real networks, such as actor networks, the WWW, and so on [1,13]. Fig. 3 shows the numerically computed and analytical degree distribution $N(k)$ with $m=3$, $N=10^4$ for different values of $\mu$, where the open circles, the stars and the squares denote cases of $\mu=1$, $\mu=0$ and $\mu=0.5$, respectively. We see that the distribution is clearly algebraic for $\mu=1$, whereas a plot on a semi-logarithmic scale indicates that the distribution for $\mu=0$ is exponential. The degree distribution for $\mu=0.5$ lies somewhere between these two cases, indicating a mixture of algebraic and exponential components in $N(k)$. We observe a good agreement between the analytical calculation and the simulation of a single realization.

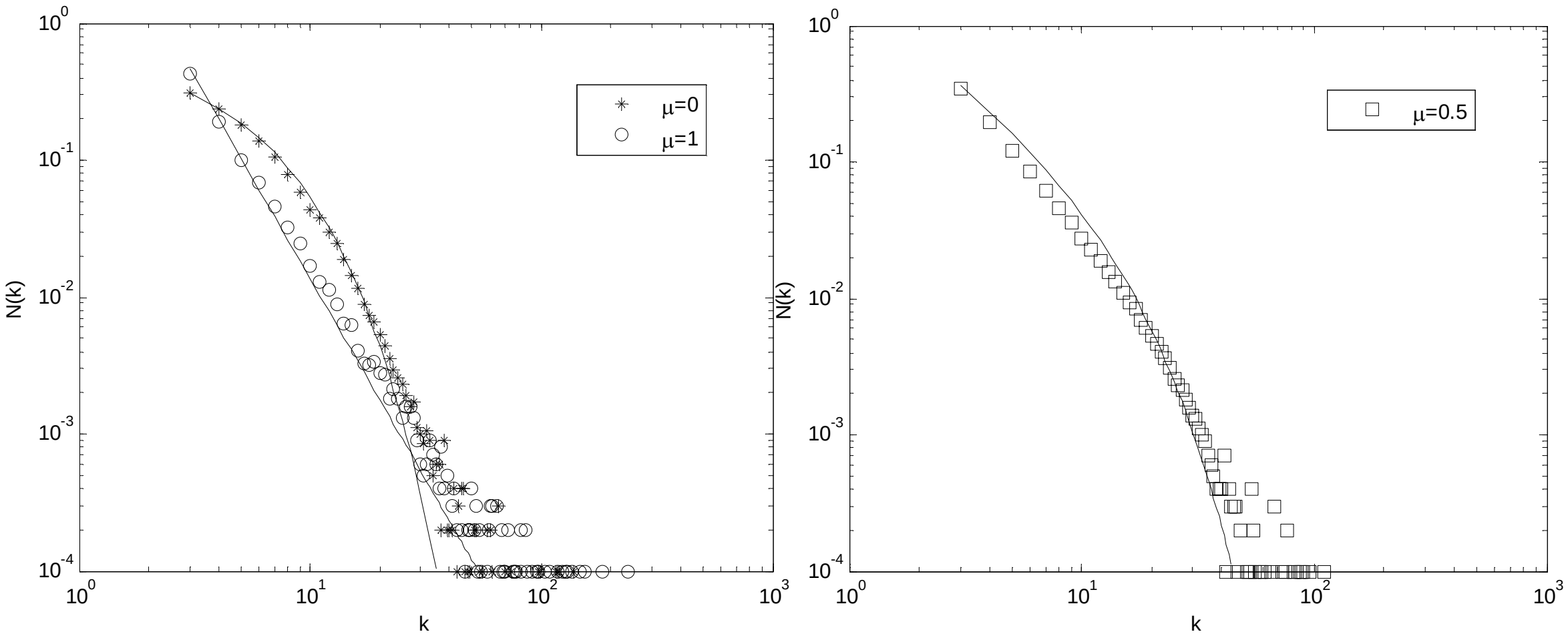

FIG. 3. Degree distribution $N(k)$ for the present model with parameter values $m=3$, $N=10^4$ at various values of $\mu$. The analytical results of degree distribution for corresponding $\mu$ are given by the solid curves with log-log scale on both axes. The numerical degree distributions by simulation are in good agreement with the analytical results.

**B. Topological self-similarity**

The emergence of topological fractal in complex networks presents the possibility of a unique growth mechanism and allows for uncovering universal origins of collective behaviors [16-19]. It springs a fundamental question for our basic understanding of the organization of complex networked systems, since the celebrated properties of small-world property and scale-free degree distribution seemed to be incompatible with self-similar features of networks [17,29]. To unfold the self-similar properties of networks, the fractal dimension is calculated using a 'box-counting' algorithm [16]. In the algorithm, the renormalization scheme titles a network of $N$ nodes with $N_B(\ell_B)$ boxes. The boxes contain nodes separated by linear size $\ell_B$, the shortest path length between nodes, and each box is successively replaced by a virtual node until the whole network is reduced to a single node. Fractal networks lead to a scaling relation $N_B / N \sim \ell_B^{-d_B}$, with an exponent that is given by the fractal dimension $d_B$.

Using 'box-counting' algorithm, it has been observed that several real networks, such as WWW ($d_B = 4.1$), actor networks ($d_B = 6.3$), protein interaction networks of *E. coli* ($d_B = 2.3$) and *H. sapiens* ($d_B = 2.3$), cellular networks of *A. fulgidus, E. coli, C. elegans* ($d_B = 3.5$), and the genetic regulatory network of two organisms *S. cerevisiae* ($d_B = 5.1$) and *E. coli* ($d_B = 3.4$), can have a fractal structure [16,29]. However, most of the random network models proposed yet are not fractal. Until very recently, Song et al. present self-similar dynamical evolution of complex networks by the inverse renormalization procedure with all of the properties of the network being invariant under time evolution. In Ref. [18], Model I produces a scale-free, small world network but without fractal topology, while Model II leads to a scale-free network with a fractal topology but not the small-world effect. They suggest growing fractal small-world and scale-free networks from an indirect way of stochastic combination of Model I and Model II. Goh et al. present an *in silico* model with both fractal scaling and scale-free degree distribution based on the multiplicative branching tree. Note also that this fractal trees are not small world, but by introducing a small number of global shortcuts yields small-world property [19]. Different from these growth mechanisms, the present model can grow fractal small-world and scale-free networks naturally.

We apply box-covering method to the present model, and the log-log plot of $N_B$ versus $\ell_B$ of the present model with different values of $\mu$ is shown in Fig. 4. In case of $\mu = 0.5$ in Fig. 4(b) we observe a power-law behavior between $N_B / N$ and $3 \le \ell_B \le 17$ with $d_B = 2.5 \pm 0.09$ . Extended calculations show perfect fractal scaling satisfied in the present model for a wide range of deactivate rate $\mu \in [0.3, 0.8]$ , with fractal dimension ranges from $d_B = 2.8$ to $d_B = 2.3$ . However, as shown in Fig. 4(a,c,d), the model exhibits a lack of fractal scaling for both $\mu \in [0, 0.3)$ and $\mu \in (0.8, 1]$ . It is noteworthy that there exists an important distinction between the present model with $\mu \in [0.3, 0.8]$ and special cases with $\mu = 1$ (power-law degree distribution) and $\mu = 0$ ( exponential degree distribution) are not fractal.

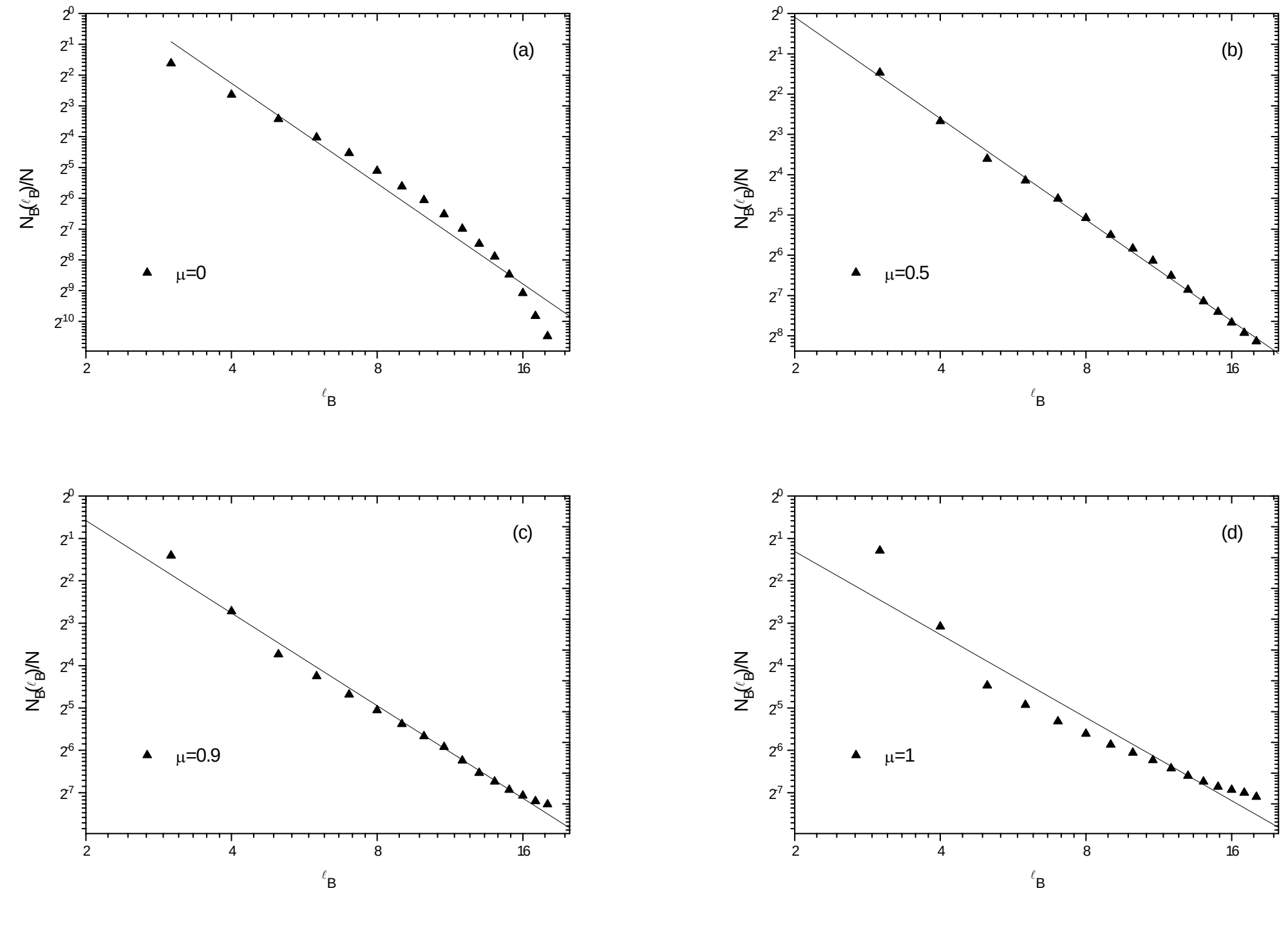

FIG. 4. Normalized number of boxes $N_B$ as a function of linear box size $\ell_B$ in the present model, where $m = 3$ , $N = 10^4$ . (a) $\mu = 0$ . (b) $\mu = 0.5$ . (c) $\mu = 0.9$ . (d) $\mu = 1$ .

## IV. SMALL-WORLD EFFECT OF THE PRESENT FRACTAL NETWORK

### A. Clustering coefficient

By definition, the clustering coefficient $C$ of a vertex is the ration of the total number of existing connections between all its $k$ nearest neighbors and the number $k(k-1)/2$ of all possible connections between them. We can go beyond the degree distribution and compute the clustering coefficient $C(k)$ as a function of the vertex degree $k$. For this quantity we can perform an analytic calculation for any value of $m$ for the provided model. In the present model, new edge is created between the active vertex and the added one, and the other $m-1$ edges are linked to its nearest neighbors. The total number of connections between all its $k$ nearest neighbors increases by $m-1$ every time when the degree $k$ increases by one. Obviously, $k_i$ and $e_i(k)$ remain constant for inactive vertices and increase only for active vertices. Therefore, the dynamics of $e_i(k)$ is given by

$$\frac{de_i(k)}{dk} = m-1 \tag{6}$$

when a new vertex $i$ is created, the degree $k_i$ of the vertex $i$ is $m$, thus $e_i(m) = m(m-1)/2$. Integrating Eq.(7) with this initial condition, we obtain

$$e_i(k) = (m-1)(k-m) + m(m-1)/2 \tag{7}$$

which gives

$$C_i(k) = \frac{2(m-1)}{k} - \frac{(m-1)(m-2)}{k(k-1)} \tag{8}$$

This expression indicates that the local clustering coefficient $C(k)$ scales as $k^{-1}$, indicating that the present networks have a hierarchical topology, which is a fundamental characteristic of many complex systems, such as th4e World Wide Web, actor network, and the Internet at domain level [15]. In Fig. 5(a), we plot the clustering coefficient as a function of the vertex degree obtained for present model, which coincides with the analytical expression in Eq.(8).

The clustering coefficient $C$ of the whole network is the average of $C(k)$ over all vertices, i.e.,

$$C = \int_m^\infty C(k)N(k)dk \tag{9}$$

For $\mu = 1$, we have $C = \frac{5}{6} - \frac{7}{30m} + \mathrm{O}(m^{-2})$. In the opposite case of $\mu = 0$, the value of clustering coefficient is $C_{m=3} = 0.66$. Generally, the analytic clustering coefficient $C$ varies between 0.66 and 0.76 for $0 < \mu < 1$ in case of $m = 3$, which is also confirmed in Fig. 6(a). The clustering coefficient of the present model is similar to the one obtained in the actor network ($C = 0.79$), the coauthorship network in spires ($C = 0.726$), and networks of word synonyms ($C = 0.7$) [1]. Fig. 5(b) shows that the average value of the clustering coefficient $C$ does not depend on the network size $N$. However, in the BA model, the clustering coefficient decreases drastically with growing system size.

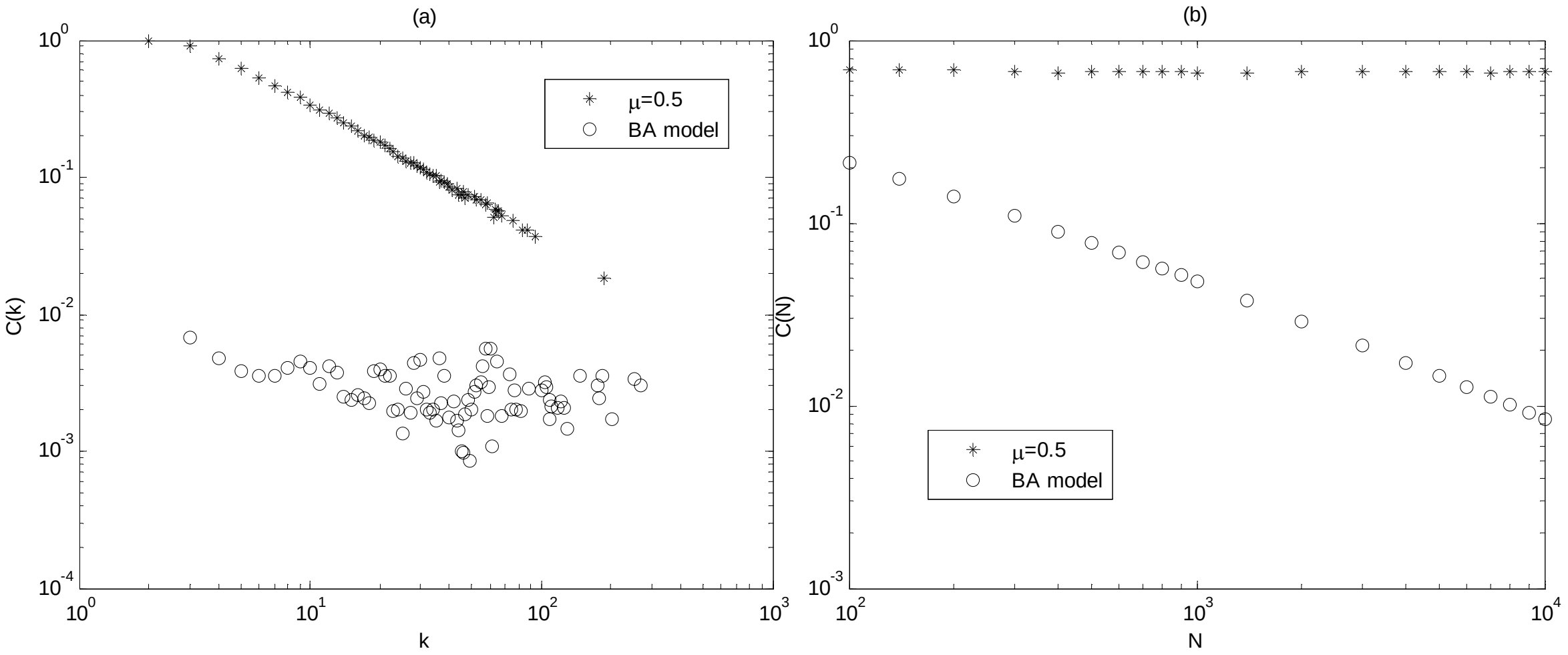

FIG. 5 Illustration of the local and average clustering coefficient with various vertex degrees $k$ and system sizes $N$. (a) The scaling of the local clustering coefficient $C(k)$ with $k$ for the present model and BA model, where $m = 3$, $N = 10^4$. (b) The clustering coefficient $C$ as a function of network size $N$. The clustering coefficient of networks generated with $\mu = 0.5$ (stars) is almost constant and independent with the network size. The clustering coefficient of BA model (circle) decreases with the increasing of network size quickly. All values plotted are averages over 100 independent realizations. The average degree is $\langle k \rangle = 6$.

**B. Characteristic path length**

Another fundamental topological feature of complex networks is identified by the scaling of the average shortest path length among vertices. In Fig. 6 we show the average shortest path length $L$ of the provided model as the functions of deactivate rate $\mu$ and the system size $N$. For $\mu = 1$, i.e., power-law degree distribution, the average shortest path length grows linearly, $L \propto N$, the same behavior observed in one-dimensional regular lattice. Ref. [30] shows that the networks' topology, for such a special case, is similar to a chain of dense clusters locally connected. Since the number of active nodes $m$ remains unchanged in growing networks, on one hand, those active nodes with long life-time have a possibility to develop a hub, and once they are deactivated, they will not receive any further links. On the other hand, the chains will grow with those bridge nodes, rapidly deactivated active nodes, until a new dense cluster is developed. The growing mechanism, without effective shortcuts that are able to reduce the path length, leads to a lack of small-world property. In contrast, the case of $\mu = 0$, i.e., exponential degree distribution, shows a slow (logarithmic) increase of the average path length of the network with the total number of nodes, $L \propto \ln N$. Ref. [28] shows why $L$ grows more slowly than $N$ although the added node links locally to the existing nodes. In such a special case, all existing nodes remain active in the growth of networks. The older nodes that have once been nearest neighbors will be pushed apart as newer nodes are inserted, thus, have a large number of newer nodes between them. Therefore, the edges that link the old nodes will server as shortcuts, responsible for a short average path length. For networks with small value of $\mu$， the initial links between the older nodes, which will remain active for a period long enough that newer nodes will be inserted successively, will more likely to be the long-range connections. On the opposite side, few shortcuts will be formed for large values of $\mu$, and the average path length increases. The characteristic path length $L(\mu)$ varying with $1-\mu$ is shown in Fig. 6(a). In case of $\mu \in [0, 0.9)$, the average path length $L$ increases slowly until an emergence of a jump when $\mu \in (0.9, 1]$, meanwhile the clustering coefficient remains almost unchanged. Interestingly, as shown in Fig. 6(b), we find an almost logarithmic growth of the average shortest path length with system size for $\mu \in [0, 0.8)$. This in addition to the high clustering yields typical of the small-world effect.

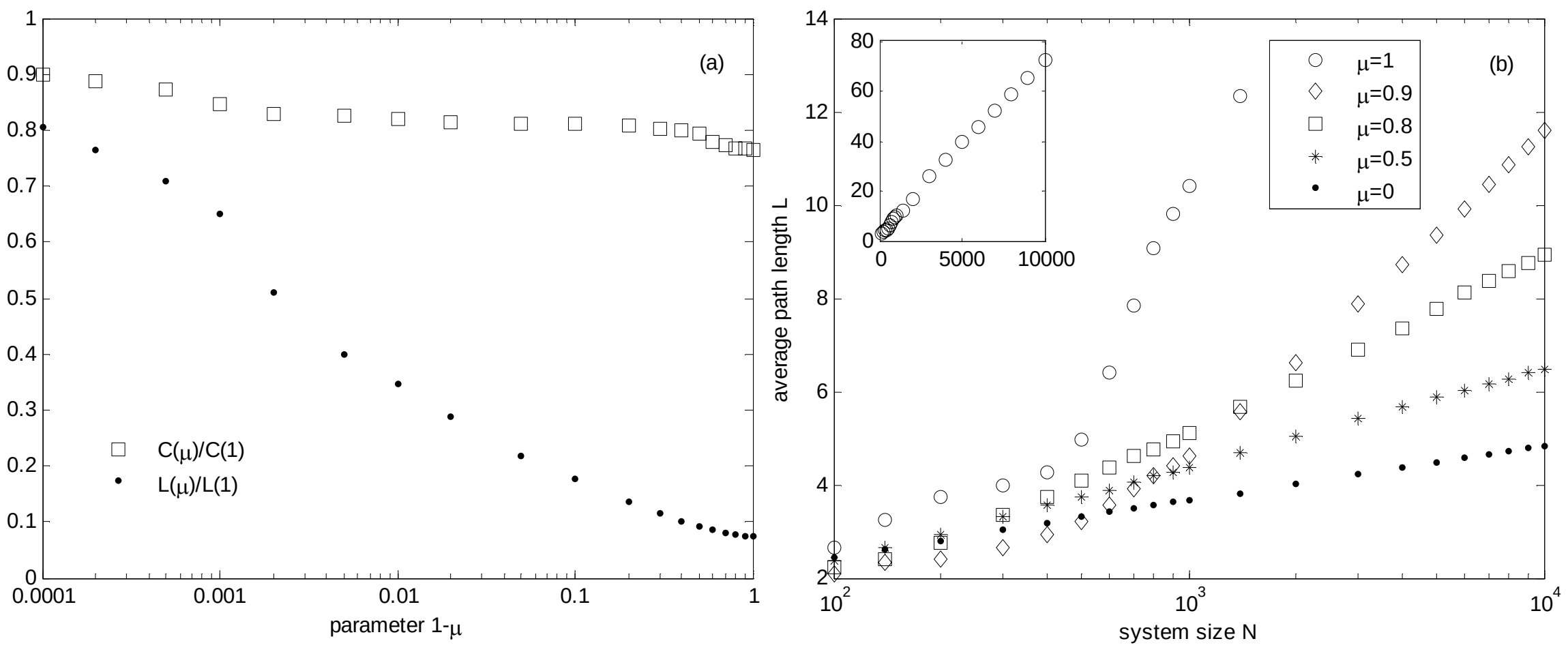

FIG. 6. Characteristic path length $L(\mu)$ and clustering coefficient $C(\mu)$ varies with $1-\mu$ and the average shortest path length $L$ as a function of system size $N$ for the family of the present model. (a) Introducing the deactivate rate $0 \le \mu \ll 1$ into the growth networks drastically reduces the distance $L$ between nodes. When $\mu$ approaches to the order of 1, the average path length increases significantly, meanwhile the average clustering coefficient $C(\mu)$ remains almost unchanged. The data shown in the figure have been normalized by the value $L(1)$ and $C(1)$, respectively. (b) When $0 \le \mu \le 0.8$, $L$ grows almost logarithmically with $N$, while $\mu > 0.8$, the exponential dependency between $L$ and $N$ weakens, and the network degenerates into KE model when $\mu = 1$, whose average path length increases linearly with system size, $L \propto N$. All plotted values are averages over 100 independent realizations and with an average degree of $\langle k \rangle = 10$.

## V. CONCLUSION

We have defined a simple model of self-organizing networks based on empirical evidence of social behaviors. The model is growing on two coupled reasonable mechanisms: the geographical attachment preference and branching structured deactivation mechanisms. We focus on the connection between the mixed degree distribution and topological self-similarity, and also analyze the structural properties such as clustering coefficient, and average shortest path length systemically. The network yields a spectrum of degree distribution ranging from algebraic to exponential and average shortest path length ranging from linear to logarithmic simply by changing a control parameter: deactivate rate, introducing the topological fractal property in a wide range of deactivate rate. In both limit cases of $\mu = 0$ (exponential degree distribution) and $\mu = 1$ (power-law degree distribution), the networks are not fractal. When $\mu$ approaches to 0, the networks feature power-law degree distribution and high clustering, but the average path length depends linearly on system size. While $\mu$ approaches to 1, the networks are characterized by small-world effect, but possess an exponential degree distribution. In general, the network yields mixed degree distribution, topological fractal structure and small-world effect in a wide range of deactivate rate. The present networks growth mechanism presents a reliable connection among small-world graphs, scale-free networks and fractal topological networks, and gives a further insight into understanding the origin of complex networks.